# Matrix Energy as a Measure of Topological Complexity of a Graph

*Kaushik Sinha[1] and Olivier L. de Weck[1]*
[1]*Engineering Systems Division, Massachusetts Institute of Technology*,
77 Massachusetts Avenue, Cambridge, MA, USA, {sinhak@mit.edu; deweck@mit.edu}

## ABSTRACT

The complexity of highly interconnected systems is rooted in the interwoven architecture defined by its connectivity structure. In this paper, we develop matrix energy of the underlying connectivity structure as a measure of topological complexity and highlight interpretations about certain global features of underlying system connectivity patterns. The proposed complexity metric is shown to satisfy the Weyuker's criteria as a measure of its validity as a formal complexity metric. We also introduce the notion of ***P*** point in the graph density space. The ***P*** point acts as a boundary between multiple connectivity regimes for finite-size graphs.

## 1 INTRODUCTION

In the context of complex interconnected systems, the quantification of complexity has increasingly gained importance over the last few years. While working with large, complex systems, the challenge of quantifying complexity is central and rigorous, formalized framework to compute and compare their respective complexities and aid decision-making.

In particular, the consideration of connectivity structure attracts attention because they affect system behavior. The term "structure" is directly linked to the definition of a system. In general, the term "structure" is understood as the network formed by dependencies between components of any system [9]. One emerging area of application is in the characterization and impact of complex system architectures that are fast becoming highly networked and distributed in nature [17, 19, 21].

The concept of network dimension can be used to determine the underlying network structure and its function. A network with higher dimension is said to be more complex than one with a lower dimension. Here, we focus on the spectral dimension of the binary adjacency matrix that represents the connectivity structure of the system. Recently, the idea of spectral dimension has been used to estimate the *reconstructability* of networks [13, 19].

In this paper, we propose *matrix energy* of the underlying binary adjacency matrix of the networked complex system as a measure of *topological complexity* of the system. We explore the properties of *matrix energy,* including important bounds for general adjacency matrices that are asymmetric. The *matrix energy* is shown to satisfy the Weyuker's criteria [20] and is therefore a mathematically valid construct for measuring complexity.

## 2 MATRIX ENERGY AND TOPOLOGICAL COMPLEXITY

Topological complexity originates from interaction between elements and depends on the combinatorial nature of such connectivity structure. The topological complexity is defined as the *matrix energy* of the adjacency matrix [6]. The adjacency matrix $A \in M_{nxn}$ of a network is defined as follows:

$$A_{ij} = \begin{cases} 1 \ \forall[(i,j) \mid (i \neq j) \text{ and } (i,j) \in \Lambda] \\ 0 \text{ otherwise} \end{cases}$$

where $\Lambda$ represents the set of connected nodes and *n* being the number of components in the system. The diagonal elements of A are zero. The associated *matrix energy* [6, 14, 16] of the network is defined as the sum of singular values of the adjacency matrix:

$$E(A) = \sum_{i=1}^{n} \sigma_i, \text{ where } \sigma_i \text{ represents i}^{th} \text{ singular value}$$

This definition extends the applicability of the metric to any simple graph, undirected and directed alike. The singular values of any matrix are always positive or zero and therefore, the matrix energy works for any simple graph. Please note that matrix energy is zero only if there are no edges and all the nodes exist in isolation. Any kind of connected graph, including purely directed graph, has a finite, non-zero matrix energy.

The *matrix energy* also expresses the minimal effective dimension embedded within the connectivity pattern represented through the binary adjacency matrix. This minimum effective dimension is expressed through the rank of the adjacency matrix and matrix energy is a convex approximation for the matrix rank [2]. Using singular value decomposition (SVD), we can express matrix A as:

$$A = \sum_{i=1}^{n} \sigma_i \underbrace{u_i v_i^T}_{E_i} = \sum_{i=1}^{n} \sigma_i E_i \qquad (1)$$

where $E_i$ represents simple, building block matrices of unit matrix energy and unit norm. Using this view, we observe that *matrix energy* express the sum of weights associated with the building block matrices required to represent or reconstruct the adjacency matrix *A*.

The matrix energy has also been used in matrix reconstruction problem, where minimization of nuclear norm was shown to yield the optimal matrix [3]. Recent research is exploring application of compressive sensing to networks [3].

## 3 PROPERTIES OF MATRIX ENERGY

### 3.1 Matrix Energy bounds for general asymmetric adjacency matrices

In case of symmetric matrix, the singular values were equal to the absolute eigenvalues and the singular vectors were directly related to signed eigenvectors. This helped us leverage some well-established mathematical properties related to the eigenvalues and establish bounds analytically. The matrix energy bounds for undirected networks can be found elsewhere [5, 7, 10, 16].

Now let us look at the extension of matrix energy bounds for mixed graphs where the links are a mix of directed and undirected ones. Here, we focus on the bounds for matrix energy for generalized asymmetric, binary adjacency matrices.

In case of a mixed graph with both, directed and undirected links, the sum of all the elements of the adjacency matrix is,

$$\| A \|_1 = \sum_{i=1}^{n} \sum_{j=1}^{n} \left| a_{ij} \right| = sm \qquad (2)$$

where $\| A \|_1$ is the Holder norm [1], defined as the sum of the absolute values of entries of the matrix and $s \in [1,2]$. Here $s = 1$ for purely directed ER graph (i.e., all links are unidirectional) and $s = 2$ for purely undirected ER graph (i.e., all links are bidirectional).

Using the Frobenius norm, $\|A\|_F$ [1, 6], we have:

$$\|A\|_F = \sum_{i=1}^{n} \sigma_i^2 = \sum_{i=1}^{n} \sum_{j=1}^{n} a_{ij}^2 = sm \qquad (3)$$

Now, we can express the squares of the sum of the singular values as,

$$\left( \sum_{i=1}^{n} \sigma_i \right)^2 = \sum_{i=1}^{n} \sigma_i^2 + 2 \sum_{1 \le i \le j \le n} \sigma_i \sigma_j$$
$$= sm + 2 \sum_{1 \le i \le j \le n} \sigma_i \sigma_j \qquad (4)$$

For the second term on the right hand side of eq. 3, while we do have a lower bound of $2m$ if A is symmetric (i.e., undirected graphs), we do not have any closed form analytical bound in case of any mixed graph where $s \in (1,2)$.

From extensive simulation studies (see fig. 1 and 2), we obtain the following lower bound,

$$2 \sum_{1 \le i \le j \le n} \sigma_i \sigma_j \ge sm \qquad (5)$$

Combing eq. 4 and eq. 5, we obtain the following lower bound for the matrix energy of general mixed graphs with asymmetric adjacency matrices,

$$\left(\sum_{i=1}^{n}\sigma_i\right)^2 \geq 2sm$$
$$\therefore E \geq \sqrt{2sm} \qquad (6)$$

Using the Cauchy-Schwarz inequality for arbitrary real-valued numbers $a_i$, $b_i$ with i =1, 2, . . . , $N$, we have

$$\left(\sum_{i=1}^{N} a_i b_i\right)^2 \leq \left(\sum_{i=1}^{N} a_i^2\right)\left(\sum_{i=1}^{N} b_i^2\right)$$

If we choose N = n, $a_i = \sigma_i$ and $b_i$ = 1, we get,

$$\left(\sum_{i=1}^{n}\sigma_i\right)^2 \leq \underbrace{\left(\sum_{i=1}^{n}\sigma_i^2\right)}_{=sm} n$$
$$\therefore E^2 \leq smn \qquad (7)$$

In order to have at least a *connected* graph, we should have $sm \geq n$, and therefore we have the following upper bound for matrix energy

$$E^2 \leq s^2 m^2$$
$$\therefore E \leq sm \qquad (8)$$

Combing eq. 5 and eq. 8, we obtain the following bounds for matrix energy of graphs with both, directed and undirected edges,

$$\sqrt{2sm} \leq E \leq sm \qquad (9)$$

Please note that for undirected graph, we have $s = 2$ and we get back the established bounds, $2\sqrt{m} \leq E \leq 2m$. For purely directed graphs, we have $s = 1$ and the bounds are given by, $\sqrt{2m} \leq E \leq m$. This bound was not found to be *tight* for random graphs.

Now, using the Cauchy-Schwarz inequality again with N = n-1, $a_i = \sigma_{i+1}$ and $b_i$ = 1, we obtain the following for mixed graphs with asymmetric adjacency matrices,

$$(E - \sigma_1)^2 \leq (n-1)(sm - \sigma_1^2)$$
$$\therefore E \leq \sigma_1 + \sqrt{(n-1)(sm - \sigma_1^2)} \qquad (10)$$

Using the earlier relations: $\sum_{i=1}^{n}\sigma^2{}_i = sm$ and $\sigma_1 \geq \frac{sm}{n}$, we have,

$$E \leq \frac{sm}{n} + \sqrt{(n-1)[sm - \left(\frac{sm}{n}\right)^2]} \qquad (11)$$

The limiting form of the above relation can be expressed in the following form for fixed *n*:

$$f(m) = sm/n + \sqrt{(n-1)[sm-(sm/n)^2]} \qquad (12)$$

Now, let us maximize the function $f(m)$, defined in eq. 12, where *n* is fixed.

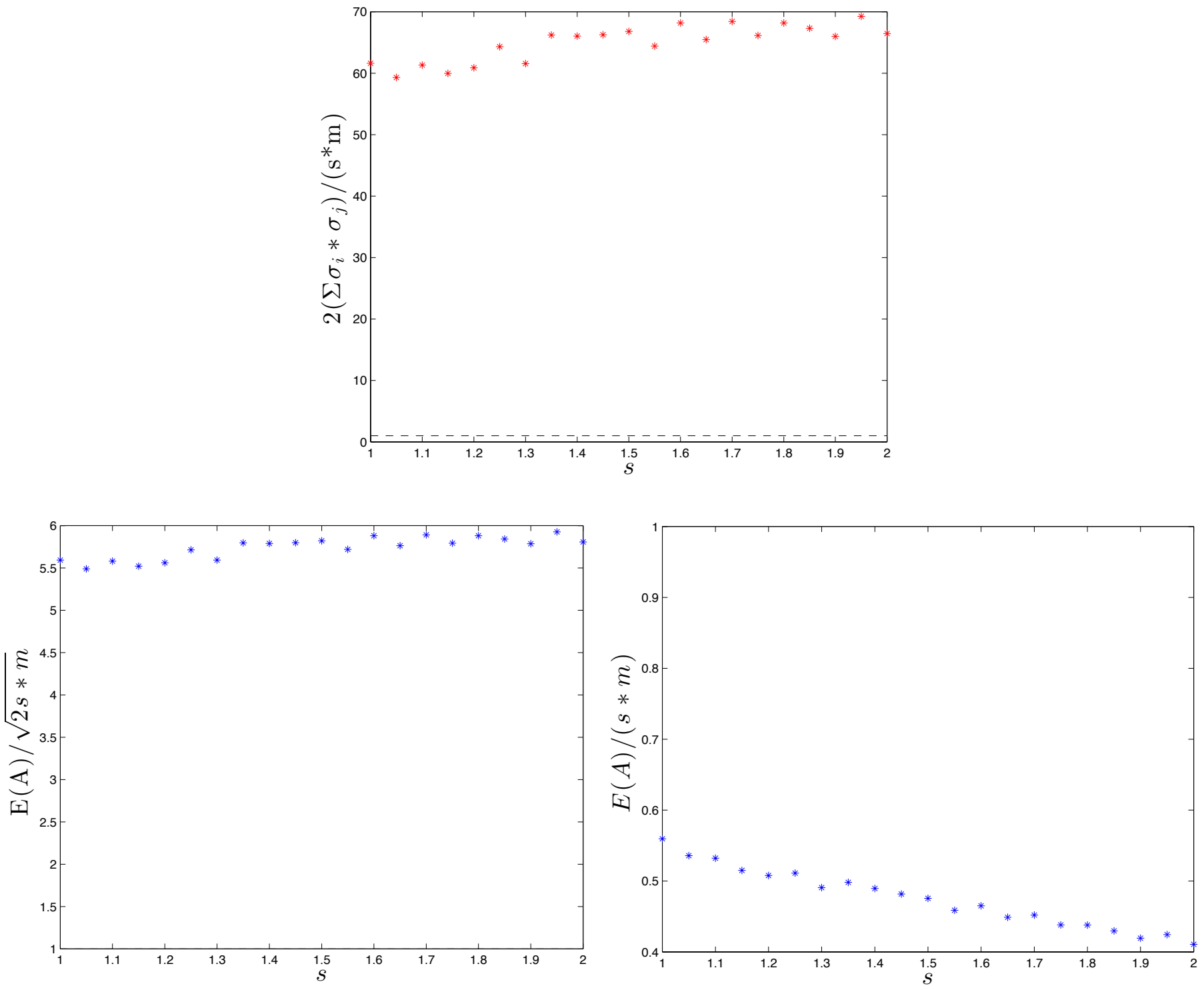

Fig. 1: Simulation results supporting (i) the bound in F.4 for mixed ER graphs with n = 100 nodes and m = 200 links; (ii) the lower and (iii) the upper bound of matrix energy, E(A) for asymmetric adjacency matrices, modeling mixed ER graphs.

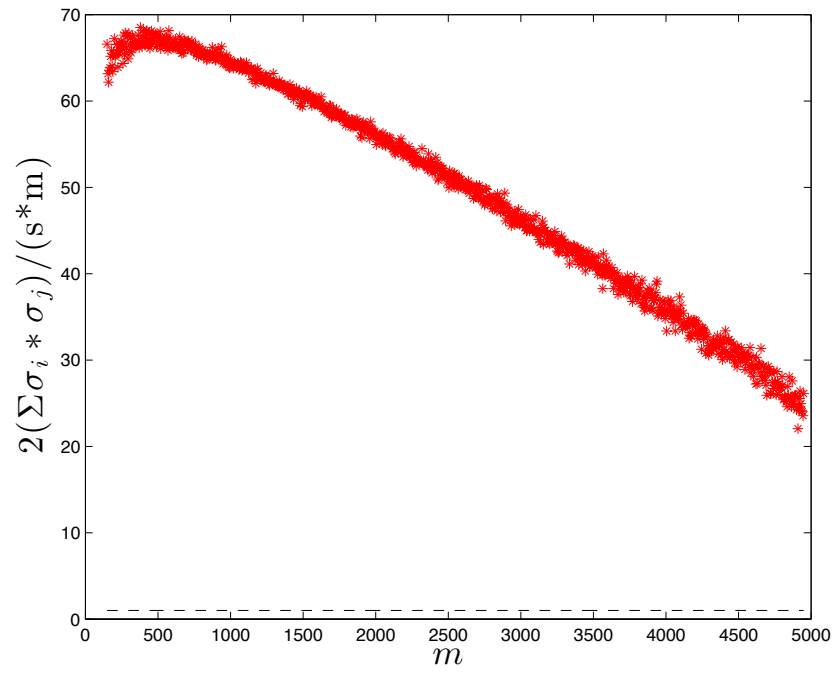

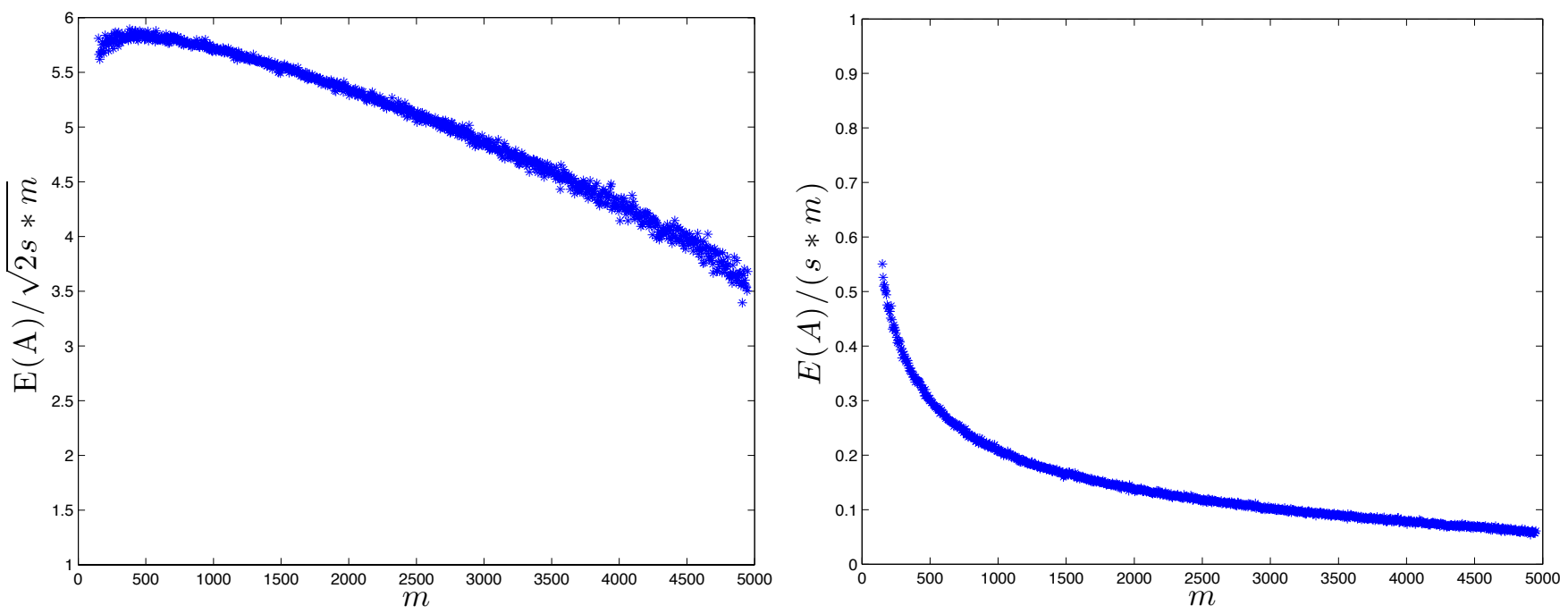

Fig. 2: Simulation results supporting (i) the bound in Eq.4 for mixed ER graphs with n = 100 nodes, s = 1.5 with varying number of links, *m*; (ii) the lower and (iii) the upper bound of matrix energy, E(A) for asymmetric adjacency matrices with n = 100 nodes, s = 1.5 with varying number of links, *m*.

Applying the *Kuhn-Tucker* optimality criteria [15], we should have:

$$\frac{df}{dm} = 0$$

$$\Rightarrow \frac{s}{n}\left\{1 + \frac{(n-1)(n^2 - 2sm)}{2\sqrt{(n-1)sm(n^2 - sm)}}\right\} = 0$$

On algebraic simplification, we get,

$$(n-1)(n^2 - 2sm)^2 = 4sm(n^2 - sm)$$

$$\therefore m = \frac{n^2 + n^{3/2}}{2s} = \frac{n^2\left(1 + \frac{1}{\sqrt{n}}\right)}{2s} \qquad (13)$$

Using the above result, we compute the corresponding value limiting matrix energy for the general case of asymmetric adjacency matrices with both, directed and undirected links,

$$f = \frac{n\left(1 + \frac{1}{\sqrt{n}}\right)}{2} + \sqrt{(n-1)\left[\frac{n^2\left(1 + \frac{1}{\sqrt{n}}\right)}{2} - \frac{n^2\left(1 + \frac{1}{\sqrt{n}}\right)^2}{4}\right]}$$

$$= \frac{n\left(1 + \sqrt{n}\right)}{2}$$

Therefore, for the maximal limiting value of matrix energy in this case, we get:

$$m^* = \frac{n^2\left(1+\frac{1}{\sqrt{n}}\right)}{2s} \approx O(n^2) \qquad (14)$$

$$f_{\max} = \frac{n(1+\sqrt{n})}{2} \approx O(n^{3/2}) \qquad (15)$$

From eq. 11 and eq. 15, we conclude that,

$$E_{\max} \le \frac{n\left(1+\sqrt{n}\right)}{2} \qquad (16)$$

Therefore, the maximal matrix energy is bounded by $n^{3/2}$,

$$E_{\max} \le n^{3/2} \qquad (17)$$

The maximal matrix energy bound is valid for generic mixed graphs, containing both directed and undirected edges or interfaces.

The results in this section extend the results derived analytically in [7, 10, 11, 16] for the special case of undirected graphs to general adjacency matrices.

### *3.2 Equi-energetic Graphs*

Two non-isomorphic graphs [5] are said to be *equi-energetic* if they have the same matrix energy. There exist finite pairs of graphs with identical spectra, called *co-spectral graphs* [5, 8]. All such co-spectral graphs are off course trivially equi-energetic. There exist finite non co-spectral graphs whose matrix or graph energy are equal and are therefore *equi-energetic*. It has been observed that there are only finite numbers of co-spectral graphs and this number tends to zero as the number of nodes, *n* increases [12]. Also, there exist finite non co-spectral, equi-energetic graphs and that number goes to zero with increasing number of nodes, *n* in the graph. Hence there could distinct graph structures with the same matrix energy and therefore, topological complexity.

## 4. *Matrix Energy as a Complexity Metric*

To check the construct validity of matrix energy as a complexity metric, let us use Weyuker's criteria [20] as the benchmark. We can think of this set of nine criteria as *necessary conditions* for theoretical or construct validity of any proposed complexity metric. In this section we show that matrix energy meets the Weyuker's criteria and qualifies as an analytically valid complexity metric construct.

Let *A* and *B* are two different systems and *K(A)* denotes complexity of system *A*:

1. ***There exist A and B such that K(A) differs from K(B),* for a complexity metric, which gives the same value for all systems is useless.**

It is well known that the number of non-isomorphic, *equienergetic* graphs (i.e.,

graphs with equal matrix energy) diminishingly small and tends to zero as we increase the graph size, *n* [12]. Hence the matrix energy differs with near certainty for different graphs.

2. ***There exist only finitely many systems of complexity c,* for the metric should have a range of values.**

As we have observed from the properties of matrix or matrix energy, there are finitely many *equi-energetic* graphs of size *n* and their number tends to zero as the graph size increases [12]. Hence, the proposed matrix energy based topological complexity metric satisfies this criterion.

3. ***There exist distinct systems A and B for which K(A) = K(B),* for a metric which gives each system unique value is not useful since such a metric would be a simple bijective mapping of systems.**

There exist finite non-cospectral, equi-energetic graphs and that number goes to zero with increasing number of nodes, *n* in the graph. Hence, it is possible to have distinct graph structures with the same matrix energy and therefore, equal topological complexity.

4. ***There exist functionally equivalent systems A and B for which K(A) ≠ K(B),* for the structure of the system determines its complexity.**

The *function to form* mapping is not unique as the same (or nearly identical) functionality can be achieved using different system architectures. They may use different *concepts* to achieve the same functionality. The differences in their system structure almost certainly yields distinct matrix energy values and therefore, have distinct topological complexities.

5. ***For all A and for all B, K(A) is smaller than K(A∪B), and K(B) is smaller than K(A∪B),* for a system is more complex than its subsystems.**

Let us define, $\Lambda = A \cup B$. Using the *pinching inequality* [6], we have,

$$E(\Lambda) \geq E(A) + E(B)$$

Hence, we have, $E(\Lambda) \geq E(A)$ and $E(\Lambda) \geq E(B)$. In case of system structures formed by *coalescence* of two graphs [12], we conclude that $E(A \circ B) \geq E(A)$ and $E(A \circ B) \geq E(B)$ since with addition of nodes while keeping the basic system structure constant leads to an increase of matrix energy [12].

6. ***There exists A, B, and M such that, K(A) = K(B) and K(A∪M) ≠ K(B∪M),* for *M* may interact with *A* in different manner than with *B.* Namely, the interface structure between *A* and *M* may be more complex than interfaces between *B* and *M*.**

Let us consider the following system structure $\Lambda_1$ where *A* and *M* represent the subsystems while $X$ represents the interfaces between the two:

$$\Lambda_1 = \begin{bmatrix} A & X^T \\ X & M \end{bmatrix}$$

The above system structure can be represented in the following block matrix form:

$$\Lambda_1 = \begin{bmatrix} A & 0 \\ 0 & 0 \end{bmatrix} + \begin{bmatrix} 0 & 0 \\ 0 & M \end{bmatrix} + \begin{bmatrix} 0 & 0 \\ X & 0 \end{bmatrix} + \begin{bmatrix} 0 & X^T \\ 0 & 0 \end{bmatrix}$$

$$\Rightarrow E(A) + E(M) \le E(\Lambda_1) \le E(A) + E(M) + 2E(X)$$

$$\Rightarrow E(\Lambda_1) = E(A) + E(M) + \Delta(X)$$

Hence, the resultant matrix energy of the integrated system has an *integrative matrix energy* component given by $\Delta(X)$.

Similarly, the system structure $\Lambda_2$ where *B* and *M* represent subsystems and *Y* represents the interfaces between the two:

$$\Lambda_2 = \begin{bmatrix} B & Y^T \\ Y & M \end{bmatrix}$$

The above system structure can be represented in the following block matrix form:

$$\Lambda_2 = \begin{bmatrix} B & 0 \\ 0 & 0 \end{bmatrix} + \begin{bmatrix} 0 & 0 \\ 0 & M \end{bmatrix} + \begin{bmatrix} 0 & 0 \\ Y & 0 \end{bmatrix} + \begin{bmatrix} 0 & Y^T \\ 0 & 0 \end{bmatrix}$$

$$\Rightarrow E(B) + E(M) \le E(\Lambda_2) \le E(B) + E(M) + 2E(Y)$$

$$\Rightarrow E(\Lambda_2) = E(B) + E(M) + \Delta(Y)$$

In this case, the resultant matrix energy of the integrated system has an *integrative matrix energy* component given by $\Delta(Y)$.

Assuming, $E(A) = E(B)$, the difference in matrix energy between $\Lambda_1$ and $\Lambda_2$ is:

$$E(\Lambda_1) - E(\Lambda_2) = \Delta(X) - \Delta(Y)$$

Hence, the difference in their matrix energy depends on how the individual subsystems interface with each other. Their matrix energies can only be equal if the subsystem interfaces are identical.

7. ***There are systems A and B such that B is a permutation of components of A and K(A)≠ K(B),*** **for changing the way connecting the components to each other, may change the level of complexity.**

The largest difference in matrix energy with same number of nodes is bounded by $\Delta E \leq \left[\frac{1}{4} + o(1)\right] n^{3/2}$ [12]. The difference stems from the way nodes of the system are connected to each other.

8. ***If A is a renaming of B, then K(A) = K(B),* for complexity does not depend on the naming of the system. This relates to the property of invariance under isomorphic transformation.**

By definition, the singular values of a matrix are independent of any rearrangement of its rows and columns. The singular values are invariant to isomorphic transformation. Therefore the matrix energy, are invariant under isomorphic transformation of the graph [6].

9. ***There exist A and B such that K(A)+K(B) is smaller than K(A∪B),* for putting systems together creates new interfaces. This pertains to the notion of "system is greater than the sum of parts".**

This criterion is satisfied due to the *pinching inequality* [6]. For a partitioned matrix

$$\Lambda = \begin{bmatrix} A & X \\ Y & B \end{bmatrix}$$

where both A and B are square matrices, we have: $E(\Lambda) \geq E(A) + E(B)$. Equality holds if and only if X and Y are all zero matrices. Hence, introduction of edges in the process of connecting two disparate graphs results in an increase in total matrix energy of the aggregated system and therefore increases the topological complexity. This inequality says that the system is more complex that its constituent sub-systems – that is, the system is larger than the sum of parts.

## 5. *P-POINT* AS A PLAUSIBLE TRANSITION POINT

The matrix energy regime for graphs with a given number of nodes can be divided into: (i) *hypo-energetic* and (ii) *hyper-energetic*. The *hyper-energetic* regime is defined by matrix energy greater than or equal to that of the fully connected, undirected graph,

$$E(A) \geq 2(n-1)$$

The *hypo-energetic* regime is defined as:

$$E(A) \leq n$$

There exists an intermediate regime between these two where the energy is higher than that of the *hypo-energetic* regime but is smaller than the *hyper-energetic* one [12].

The different connectivity regimes can be expressed using a normalized version of matrix energy $C_3$ as follows:

$$C_3 \equiv \frac{E(A)}{n} = \begin{cases} \geq 2\left(1-\frac{1}{n}\right) \approx 2 & \text{- \textit{hyperenergetic}} \\ <1 & \text{- \textit{hypoenergetic}} \end{cases} \quad (18)$$

There is a *transitional* connectivity regime that lies between the *hyper* and *hypo-energetic* regimes and is defined by, $C_3 \in [1,2]$.

There is a point beyond which the graph becomes hyper-energetic and we define this point as the ***P*** *point*.

Interestingly, ***P*** point also coincides the random graph becoming rank-sufficient (i.e., full-rank) on average (see fig. 3 below). We define a metric termed *Rank Sufficiency Factor* (RSF) as the normalized matrix rank,

$$RSF = \frac{r}{n}$$

where $r$ stands for the rank of the adjacency matrix. Notice that, on average, we have $RSF \approx 1$ as we enter into the hyper-energetic regime (with $C_3 \approx 2$).

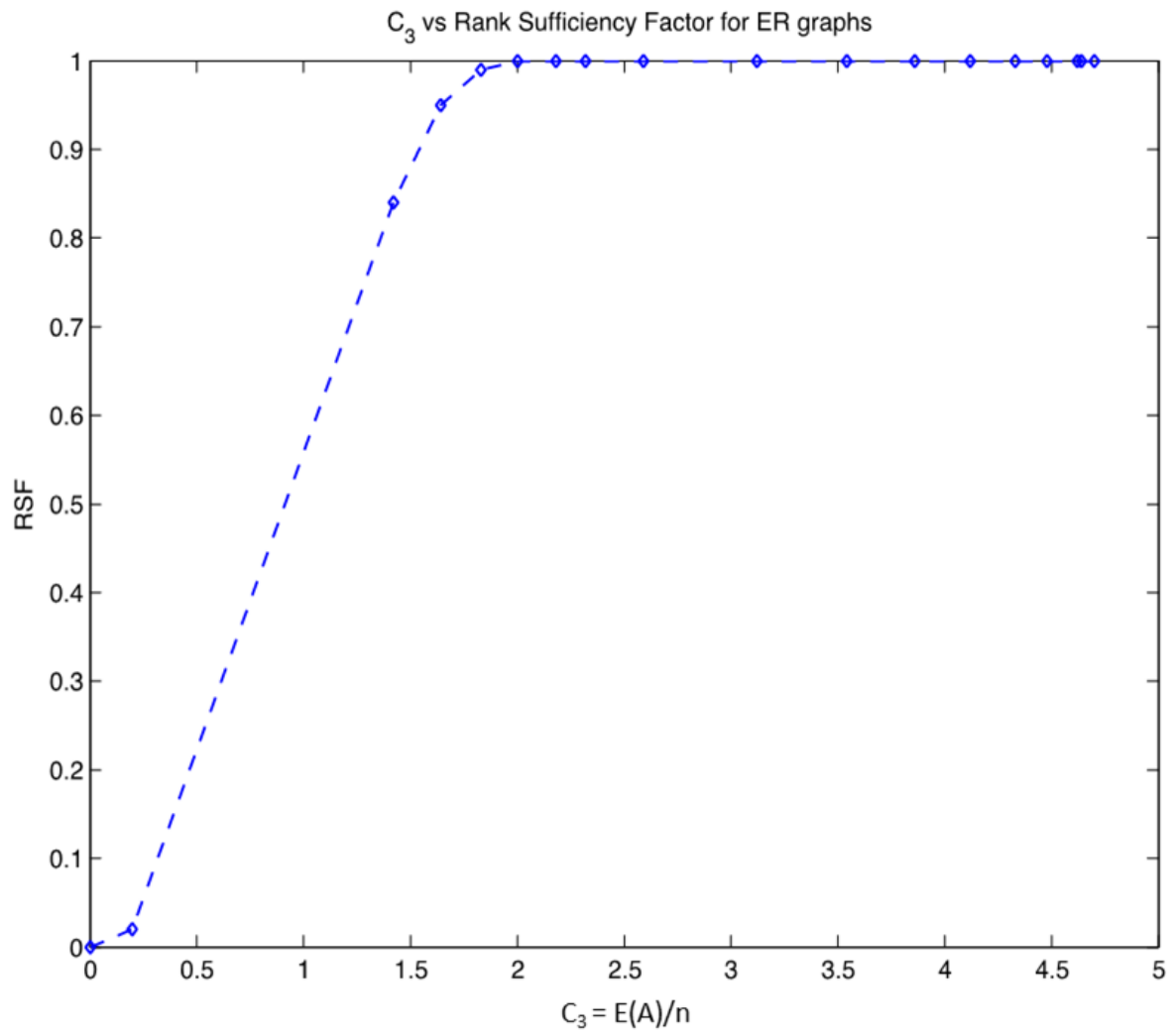

Fig. 3: Variation of $C_3$ and Rank Sufficiency Factor (RSF) for Erdos-Renyi random graphs with n = 100 nodes. The adjacency matrix becomes rank-sufficient (i.e., full-rank) around $C_3 \approx 2$, and this point is defined as the *P* point. At this point, the graph energy, *E(A)* becomes equal to that of the fully connected graph.

The ***P*** point is therefore characterized by $E(A)/n \approx 2$ and $rank(A)/n \approx 1$. All graphs on the right of ***P*** point are hyper-energetic.

Let the graph density corresponding to point *P* be termed the *critical density*, $\mu_{cr}$ and critical average degree, $\langle k \rangle_{cr}$ be the corresponding average degree of the graph. They are related as:

$$\mu_{cr} = \frac{\langle k \rangle_{cr}}{n-1} \qquad (19)$$

Above this density, the corresponding graph becomes *hyper-energetic*. The average degree of *non hyper-energetic* networks is less than this critical value of average degree of network of given number of vertices.

After tedious algebraic manipulation, from eq. 11, we can derive [16]:

$$\langle k \rangle_{cr} \geq 4(1 - \frac{1}{n})$$

$$\therefore \quad \mu_{cr} \geq \frac{4}{n} \qquad (20)$$

Therefore, at the ***P*** point, the corresponding number of links, $m_{cr}$ is given as:

$$m_{cr} = \mu_{cr} \frac{n(n-1)}{2}$$

$$\geq 2(n-1) \qquad (21)$$

Note that, the number of edges at ***P*** point, where the graph attains rank sufficiency, is at least twice the number of edges required to attain a connected graph (i.e., the number of edges in Minimum Spanning Tree is $(n-1)$ where $n$ is the number of nodes).

Based on extensive simulation studies, averaged over 10,000 instances at each network density level, on *Erdos-Renyi* random graphs with given number of vertices, the variation of the critical density (as percentage) and corresponding critical average degree for varying the graph size is shown in fig. 4 below. We can observe that as graph size increase, the critical density reduces, but the critical average degree tends to remain almost constant around a value, $\langle k \rangle_{cr} = 6$.

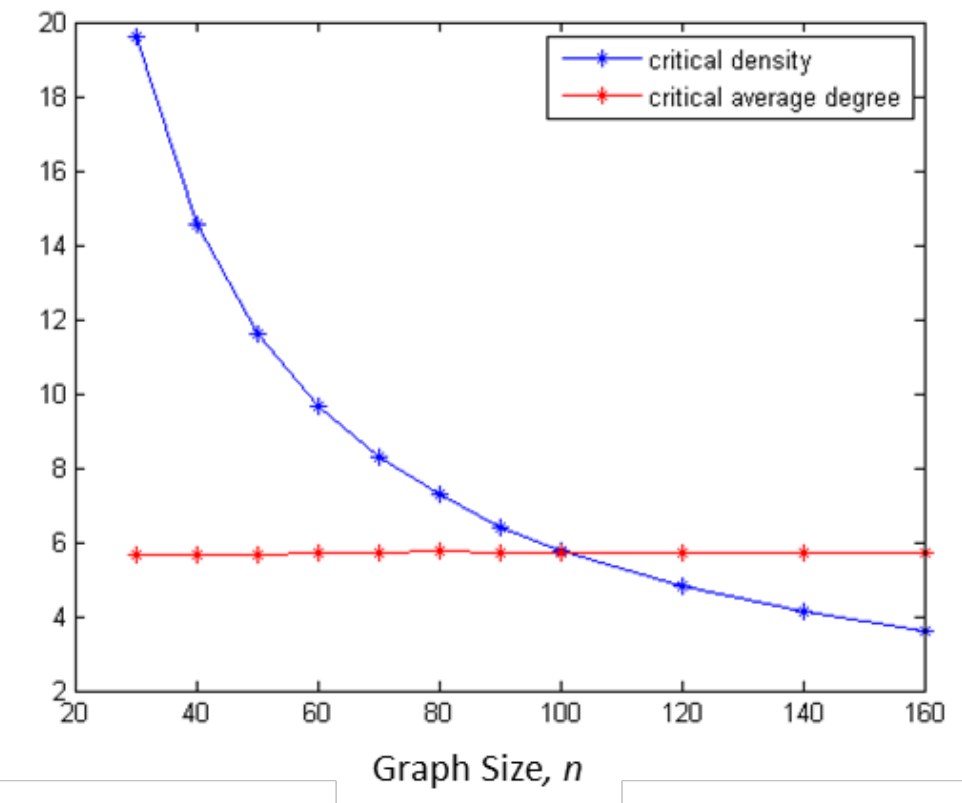

Fig. 4: Variation of critical density (%) and corresponding average degree with increasing graph size.

The ***P*** point shows very interesting features relating to interesting characteristics of the graph. It appears that nearness to rank-sufficiency of the network has important bearing on other network metrics as well. This is the density at which a network attains rank-sufficiency and becomes distributed in nature. It was observed that such distributed characteristics favor robustness against attacks and failures.

Simulation results indicate saturation in terms of relative improvement in other network metrics like maximum diameter, average path length and mixing time over networks [4, 16].

Another interesting observation was made in an analytical study by [18] regarding the resilience of general random networks against both, targeted and random attack on nodes. They defined two metrics to measure the resilience against nodal failures:

$f_a$: fraction of targeted nodes before the giant component vanishes.
$f_r$: fraction of randomly deleted nodes before giant component vanishes.

They analyzed the ($f_a$ vs. $f_r$) envelope for general random graphs. The envelopes are shown in fig. 5 (b) and it appears that the outward growth of the envelope saturates beyond the average degree, <k> = 6 (see fig. 5 below) and coincides with ***P*** point on the graph density plot.

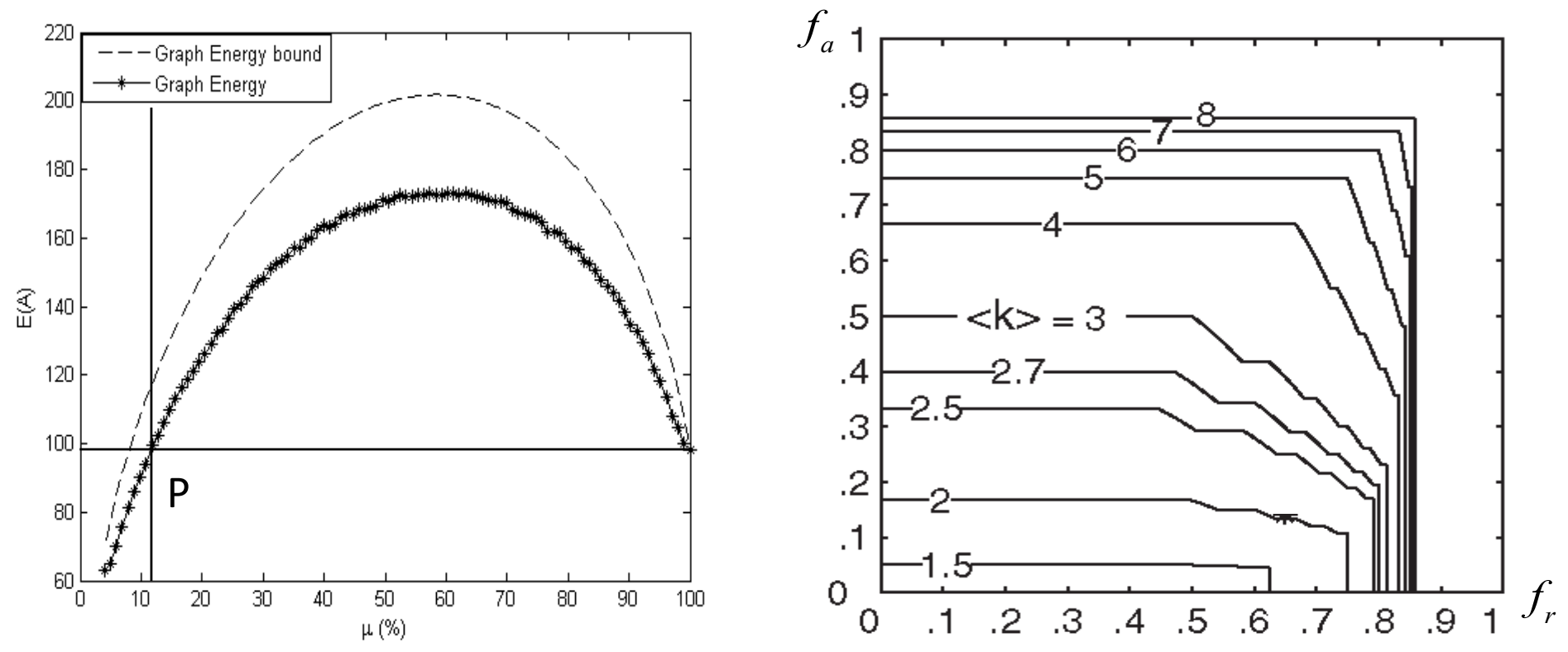

(a) (b)

Fig. 5: Characterization of (a) the *P* point on graph energy vs. graph density plot, and (b) the network resilience contour ($f_a$ vs. $f_r$) for general random graphs [Valente *et al.*, 2004] – beyond the average degree <k> = 6, the is minimal outward growth of the network resilience envelope.

In an empirical study using a large and diverse set of engineered products and systems [21] showed that the average number of connections to any component (i.e., nodes) was about 6. It appears that the ***P*** point might suggest an important system architecting guideline. Please note that this is primarily a simulation based finding at this point and requires more theoretical and empirical work in future.

## 6. DISCUSSIONS

In this work, we introduce *matrix energy* as a measure of topological complexity of networks. Topological complexity is a global measure that encapsulates the inherent structure in the system structure that encapsulates the inherent arrangement of connections. Some important properties and bounds of matrix energy for mixed graphs, with both directed and undirected edges were explored. The maximum value of matrix energy was found to be $O(n^{1.5})$. It was shown that matrix energy satisfies the Weyuker's criteria and therefore qualifies as a mathematically valid complexity metric. An empirical validation was demonstrated in a recent work [17] using simple experiments with human subjects

The behaviors of matrix energy with increasing network density were investigated and the notion of ***P*** point was introduced. The average degree of finite graphs at ***P*** point was observed to be invariant of network size and found to have a constant value of about ***6*** for random networks of finite size.

Going forward, we believe that the matrix energy can serve as a measure of topological complexity for highly networked and distributed systems.

## ACKNOWLEDGEMENT

This research was supported by Pratt & Whitney (MIT project number 6921437), the Defense Advanced Research and Projects Agency (DARPA) and Vanderbilt University (VU-DSR #21807-S8).